\documentclass[conference,a4paper]{IEEEtran}
\usepackage[OT1]{fontenc} 
\usepackage{float}
\usepackage{array}
\usepackage{bm,tikz}
\usepackage{fancyhdr}
\usepackage{enumerate}
\usepackage{amsmath}
\usepackage{graphicx}
\usepackage{booktabs}
\usepackage{amsfonts}
\usepackage{verbatim}
\usepackage{multirow}
\usepackage{graphicx}
\usepackage{epstopdf}
\usepackage{cite}

\ifCLASSINFOpdf
\else
\fi

\begin{document}

\title{Blockage Modeling for Inter-layer UAVs Communications in Urban Environments}

\author{\IEEEauthorblockN{Zhi Yang$^1$, Lai Zhou$^2$, Guangyue Zhao$^1$, Shidong Zhou$^1$}
\IEEEauthorblockA{
{$^1$}
Department of Electronic Engineering, Tsinghua University\\
{$^2$}
Department of Engineering Physics, Tsinghua University\\
Beijing, China, 100084\\
Email: yang-z15@mails.tsinghua.edu.cn}}

\IEEEoverridecommandlockouts

\maketitle

\begin{abstract}
The impact of buildings blockage on the UAVs communications in different UAVs heights is studied, when UAVs fly in a given rectangular urban area. This paper analyzes the properties of blockage behavior of communication links between UAVs from layers of different heights, including two main stochastic properties, namely the blockage (or LOS) probabilities, and the state transition models (birth-death process of LOS propagation). Based on stochastic geometry methods, the relation between LOS probability and the parameters of the region and the buildings is derived and verified with simulations. Furthermore, based on the simulations of the channel state transition process, we also find the birth-death process can be modeled as a distance continuous Markov process, and the transition rates are also extracted with their relation with the height of layers.

\end{abstract}
\IEEEpeerreviewmaketitle

\section{Introduction}
With the development of technology, Unmanned Aerial Vehicles (UAVs) have been attracting more attention in recent years. Various channel models for UAV communications will be needed for system design. Large scale fading properties should be of the first importance, including path loss and shadowing. In the urban area, blockage usually happens due to high buildings, in particular for UAVs flying within relatively low height. Since such blockage leads to Non-Line-Of-Sight (NLOS) propagation scenario and additional fading loss, the stochastic property of blockage is very important for channel modeling and system design \cite{AlHourani:2014bb,Thornburg:2016hl,Bai:2014hf,Bai:2015cy}.

For some specific applications and scenarios, the flight of UAVs may be restricted to a particular area at different layers, for example, surveillance service, goods transportation, etc.. As depicted in Fig. 1, only some links will be with good property (line-of-sight propagation) and the others are blocked by the buildings, which is also regarded as blockage. Besides, while the UAV flies across this area, the link property between different layers may be changing with the change of the UAV locations \cite{Chen:2017hh}. Therefore, the probability of Line-Of-Sight (LOS) links and the birth-death process for link property should be considered.

\begin{figure}\centering
  \includegraphics[width=3in]{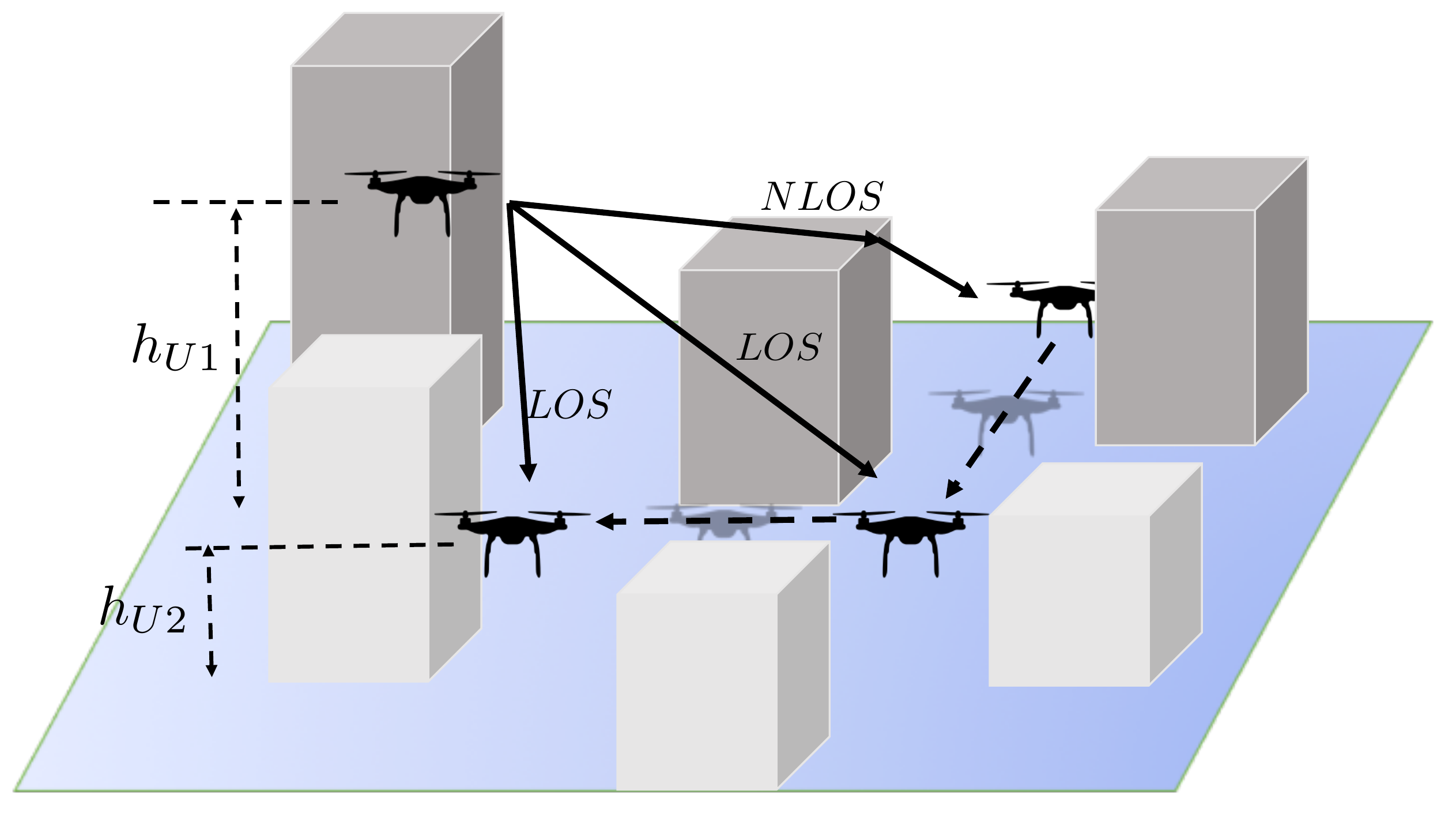}
  \vspace{-3mm}
  \caption{UAVs communication scenarios.}
  \label{fig:fig1}
  \vspace{-3mm}
\end{figure}

There are some existing related works for blockage study. International Telecommunication Union (ITU) model \cite{Baek:2015cz} expresses the LOS probability as a function of propagation distance, without considering the impact of terminal height. \cite{Bai:2014hf,AlHourani:2014gj,Holis:2008jh} propose the models associated with the elevation angle and Base Station (BS) height. However, none of these models consider the LOS probability for Air-to-Air (AA) channel in a certain area. As for the time-varying scenarios, \cite{Sato:1998bj,Kashiwagi:2010cn,Zwick:2002cq} analyze the effect of human activity on LOS state in the indoor environment, \cite{LiuLiu:2012wk}\cite{Hu:2015fr} analyze the birth-death process in the high-speed railway scenario and the vehicular networks scenario. However, there is a lack of research on low-altitude UAVs in the urban environments, where the buildings sizes and heights both affect the LOS state. In this paper, we propose the average LOS probability model for the AA channel in the urban scenario, and analyze the transition probability between LOS state and NLOS state for the Markov model. The modeling parameters are derived based on the ray-tracing simulation result.

The remainder of the paper is organized as follows: Section II introduces the models about LOS probability in urban environments, and the life distance of LOS state and NLOS state. In Section III, the stochastic geometric method is used to derive the expression of the LOS probability. In Section IV, we compare the theoretical result with the simulation result of LOS probability, and obtain the parameters of the Markov model. Section VI concludes the paper.

\section{System model}

\begin{figure}\centering
  \includegraphics[width=3in]{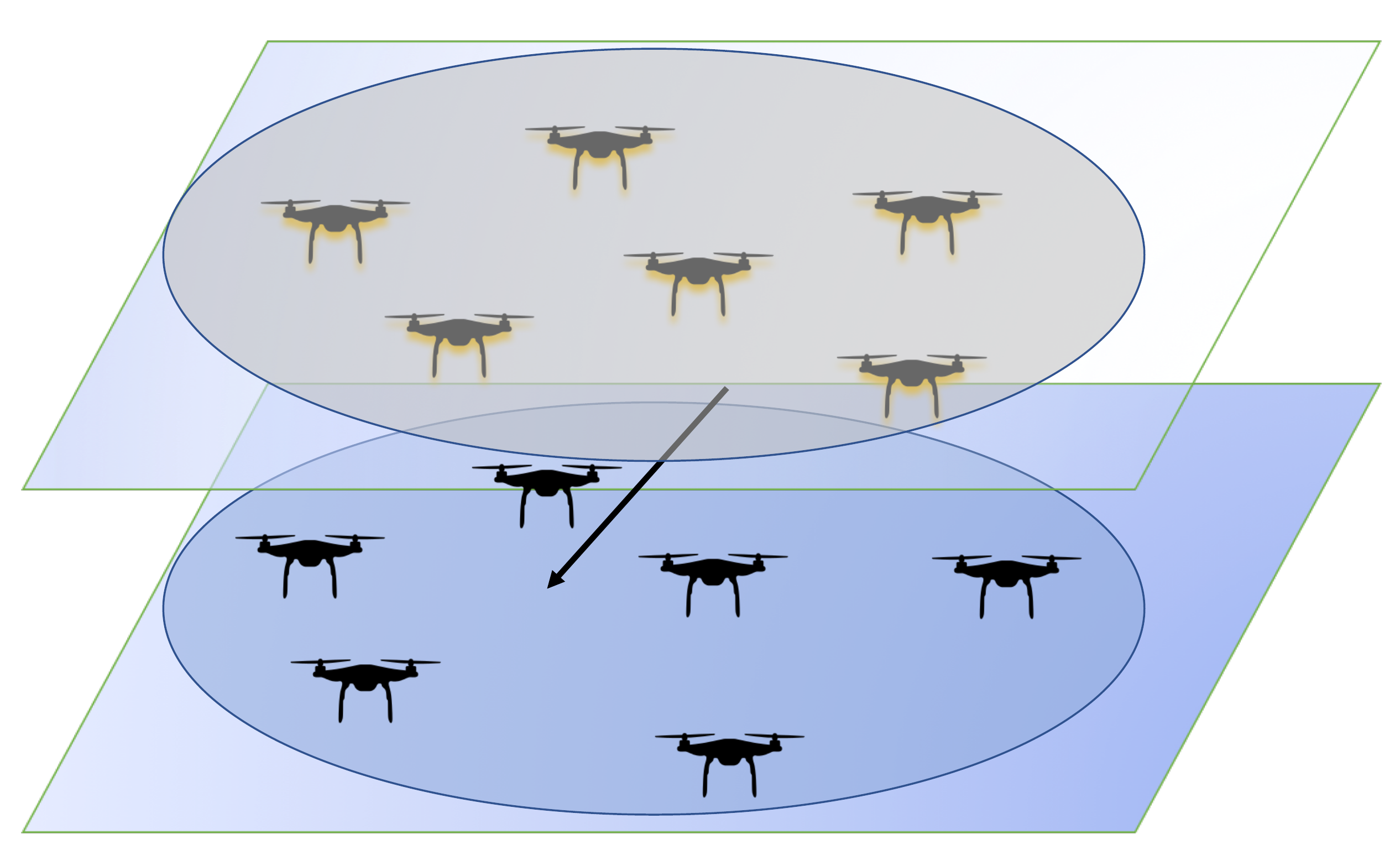}
  \vspace{-3mm}
  \caption{Multi-layers UAVs communication scenarios.}
  \label{fig:fig layer}
  \vspace{-3mm}
\end{figure}

In our paper, the UAVs at the same layer are defined as a cluster of UAVs at the same height. As depicted in Fig. \ref{fig:fig layer}, the Multi-layers UAVs are deployed in a certain area. Considering the relationship between the distribution of UAV locations and the behavior of the flights, we assumes that the UAVs are uniformly distributed in a rectangular region, and the LOS probability and the life distance of LOS/NLOS state are analyzed as below.

According to the ITU model for cellular network, the parameters will be used for building deployment. The built-up area is generated based on three simple parameters $\alpha, \beta, \gamma$ \cite{ITU}:
\begin{itemize}
  \item $\alpha$: the ratio of land area covered by buildings to total land area (dimensionless);
  \item $\beta$: the mean number of buildings per unit area ($buildings/km^2$);
  \item $\gamma$: a variable determining the building height distribution according to Rayleigh probability density function (PDF):
  \begin{equation}
    P(h) = \frac{h}{\gamma^2}e^{-\frac{h^2}{2\gamma^2}},
  \end{equation}
\end{itemize}
where $h$ is the building height in meters.

\subsection{Blockage model}
In this model, the UAV communicates with the adjacent one in the same rectangular area. The city is divided into many patches and the UAVs fly in one patch of them. The LOS probability $P_{LOS}$ is defined as the occurrence probability of a LOS link among the inter-layer UAVs communication links. Hence, the LOS probability could be expressed as:
\begin{equation}
  P_{LOS}(h_{U1}, h_{U2}) = \int_{0}^\infty f(h_{U1}, h_{U2}, l)p(l)dl,
\end{equation}
where $f(h_{U1}, h_{U2}, l)$ represents the LOS probability between the Transmitter (Tx) and Receiver (Rx) when the horizontal distance of the link is $l$,  and the heights of Tx and Rx are $h_{U1}$ and  $h_{U2}$, respectively. $p(l)$ represents the probability of $l$ when the UAVs fly in a rectangular area.

\subsection{State transition model }

\begin{figure}
  \centering
  \begin{tikzpicture}[scale=1.5]
  \path (1,0.9) node(text1)[below]{$p_{01}(d)$};
  \path (1,-0.5) node(text1)[below]{$p_{10}(d)$};
  \path (-0.8,0) node(text1)[left]{$p_{00}(d)$};
  \path (2.8,0) node(text1)[right]{$p_{11}(d)$};
  \tikzstyle{every node}=[draw,shape=circle, minimum size=15mm];
  \path (0:0cm) node (v0) {NLOS};
  \path (0:2cm) node (v1) {LOS};
  \draw [->] (v0) .. controls (0.5,0.6) and (1.5,0.6) .. (v1);
  \draw [->] (v1) .. controls (1.5,-0.6) and (0.5,-0.6) .. (v0);
  \draw [->] (v0) .. controls (-1,1) and (-1,-1) .. (v0);
  \draw [->] (v1) .. controls (3,1) and (3,-1) .. (v1);

  \end{tikzpicture}\\
  \caption{The first-order Markov model.}
  \label{fig:markov}
  \end{figure}
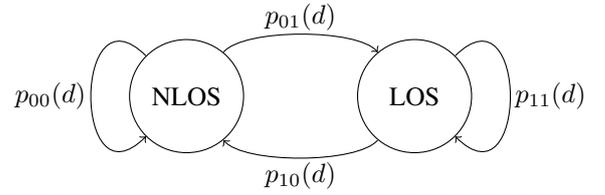

For simplicity, we analyze the transition of the LOS state when one of the UAVs is flying in the rectangular area, to build the communication link with another UAV hovering at a certain fixed altitude. As depicted in Fig. \ref{fig:markov}, the LOS/NLOS state is modeled as a distance homogeneous Markov model \cite{Dehnie:2007cm}, and there are two states, i.e., $state$ 0 (NLOS) and $state$ 1 (LOS).
A first-order Markov chain is expressed by its transition probabilities as
\begin{equation}
  p(d) = \{p_{ij}\} = \left(
  \begin{aligned}
    p_{00} \quad p_{01}\\
    p_{10} \quad p_{11}
  \end{aligned}
  \right),
\end{equation}
 where $d$ is the distance of UAV movement. $i$, $j$ represents the state index, and $p_{ij}$ is the transition probability from state $i$ to state $j$. $p_{ij}$ must satisfy the follow requirements as
\begin{equation}
  \begin{split}
    0\leq p_{ij}\leq 1, i, j\in \{0, 1\}, \\
    p_{i0}+p_{i1} = 1, i \in \{0, 1\}.
  \end{split}
\end{equation}

Assuming that the transition rate from NLOS state to LOS state is $\mu$, and the transition rate from LOS state to NLOS state is $\lambda$. Based on the Markov property, the probabilities of $p_{01}$ and $p_{10}$ are:
\begin{equation}
  p_{01}(d) = \int_0^d \mu e^{-\mu x} dx, \quad P_{10}(d) = \int_0^d \lambda e^{-\lambda x}dx.
\end{equation}
Hence, if the distance is $\Delta d$, the transition rates can be expressed as
\begin{equation}
\begin{split}
    \mu = -\frac{\ln(1-p_{01}(\Delta d))}{\Delta d},\\
  \lambda  = -\frac{\ln(1-p_{10}(\Delta d))}{\Delta d}.
  \end{split}
\end{equation}
And the expectation of the life distance $d_{LOS}$ and $d_{NLOS}$ in LOS/NLOS state can be expressed as
\begin{equation}
  E[d_{LOS}] =  \frac{1}{\lambda}, \quad E[d_{NLOS}] =  \frac{1}{\mu}.
\end{equation}
When the transition probabilities $ p_{10}, p_{01} \to 0$,
  \begin{equation}
  E[d_{LOS}] = \frac{\Delta d}{p_{10}(\Delta d)}, \quad E[d_{NLOS}] =  \frac{\Delta d}{p_{01}(\Delta d)}.
  \end{equation}
Moreover, the life time of the LOS state can be obtained as $T_{life} = f_v(d)$, where $v$ is the speed of UAV, and $f_v$ is the transform function between distance and time. The expectation of the life time in LOS state can be expressed as
\begin{equation}
  E[T_{life}] = \int_{0}^{\infty} f_v(x E[d_{LOS}])e^{-x}dx.
\end{equation}

\section{LOS probability analysis}

When there is one building, and the building height is in Rayleigh distribution as equation (1), it may affect the direct path between Tx and Rx. According to Theorem 3 in \cite{Bai:2014hf}, the LOS probability for one building is expressed as,
\begin{equation}
	P_{LOS}(1) =1-\int_{0}^{1}\int_{0}^{sh_{U1}+(1-s)h_{U2}}P(h)dh ds,
\end{equation}
 where $1$ represents one building. Let $\Delta h = |h_{U1} - h_{U2}|, h = \max(h_{U1}, h_{U2})$. Hence,
\begin{align}
	P_{LOS}(1) = \left\{
 \begin{matrix}
    1 -e^{-\frac{h^2}{2\gamma^2}} & \Delta h = 0\\
    1-\frac{\sqrt{2\pi}\gamma}{\Delta h}[Q(\frac{h}{\gamma})-Q(\frac{h+\Delta h}{\gamma})] & \Delta h > 0
  \end{matrix}
  \right. ,
\end{align}
where $Q(x) = \int_x^\infty\frac{1}{\sqrt{2\pi}}e^{\frac{-x^2}{2}}dx$. In order to estimate the effect of multiple buildings on the LOS probability, the number of buildings $N$ between Tx and Rx should be considered. In urban environments, we assume that the number of buildings between Tx and Rx at a horizontal distance of $l$ is $N \approx E[N|l]$. According to \cite{Bai:2014hf}, the expectation of the number of buildings can be expressed as
\begin{equation}
  E[N|l] = 2\sqrt{\frac{\alpha\beta}{\pi}}l+\alpha.
\end{equation}
If the impact of each building on the LOS probability is independent, the LOS probability between Tx and Rx is \cite{ITU}
\begin{equation}
  P_{LOS}(l) \approx P_{LOS}(1)^{E[N|l]}.
\end{equation}
Considering that the LOS probability is independent on the distance between Tx and Rx when they are in the same street, the corrected probability expression is:
\begin{equation}
  f(h_{U1}, h_{U2}, l) = P_0 +(1-P_0)P_{LOS}(l),
\end{equation}
where $P_0$ is the probability that Tx and Rx are located in the same street, depending on the building density. When the projection of each building on the ground is considered as a square with the same size and the building is evenly distributed in the city, $P_0$ can be expressed as
\begin{equation}
  P_0 \approx  \frac{2(1-\sqrt{\alpha})^2}{(1-\alpha)\sqrt{\beta S}}+D,
\end{equation}
where $S$ is the area of a patch in the city, and $D$ is the corrected parameter about $P_0$. Noting that $D$ is the estimated deviation due to the shape, size and distribution of the actual building, and here $D = 0.05$.

\begin{figure}\centering
  \includegraphics[width=2in]{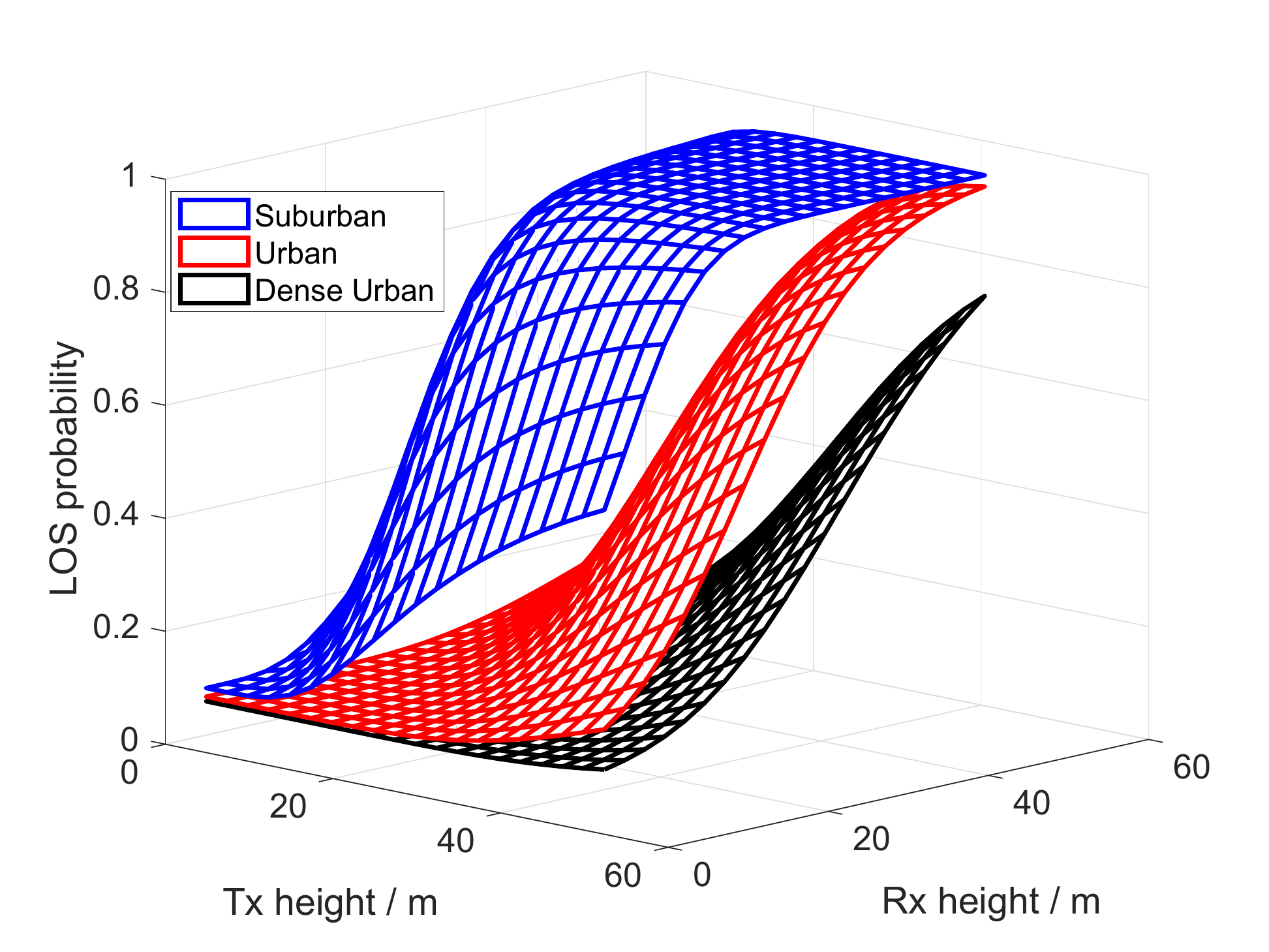}
  \vspace{-3mm}
  \caption{LOS probability in different environments, $A=775$ m.}
  \label{fig:fig LOS}
  \vspace{-3mm}
\end{figure}

The patch is assumed to be a square, and the side length is $A=\sqrt{S}$. The locations of Tx $(x_T, y_T)$ and Rx $(x_R, y_R)$ are within the area and follow a uniform distribution. Let $\Delta x = |x_T-x_R|$ and $ \Delta y = |y_T - y_R| $, the PDF of $\Delta x$ and $\Delta y$ are
\begin{equation}
  P(\Delta x = k) = P(\Delta y = k) = \frac{2k}{A^2} \quad(0\leq k\leq A).
\end{equation}
Hence, the probability of distance $l$ between Tx and Rx is
\begin{equation}
  \begin{split}
    P(\frac{l}{A}=k) = \int\limits_{\sqrt{\Delta x^2+\Delta y^2 }=kA} P(\Delta x)P(\Delta y) d\Delta x d\Delta y \\
  \approx \left\{
    \begin{matrix}
      &2\pi k - 8k^2 +2k^3&(0<k<1)\\
      & 0 & others
    \end{matrix}
    \right . ,
  \end{split}
\end{equation}
where the expectation of $l$ is $\mu = 0.52A$, and the variance of standard deviation of $l$ is $\sigma = 0.06A$. Using Gaussian distribution to approximate this expression (17) as
\begin{equation}
  p(l = k) \approx \frac{1}{\sqrt{2\pi}\sigma A}e^{-\frac{(k-\mu )^2}{2\sigma^2 }} (k>0),
\end{equation}
and taking (13)(18) into expression (2), the LOS probability is
\begin{equation}
  \begin{split}
    P_{LOS}  &=  \int_0^\infty [P_0 +(1-P_0)P_{LOS}(l)]\frac{1}{\sqrt{2\pi}\sigma }e^{-\frac{(l-\mu )^2}{2\sigma^2 }} dl \\
    &= P_0 +(1-P_0)P_{LOS}(1)^\eta \int_0^\infty \frac{1}{\sqrt{2\pi}\sigma }e^{-\frac{(l-\Gamma )^2}{2\sigma^2 }} dl
  \end{split}
\end{equation}
where $\Gamma = \mu+2\sqrt{\frac{\alpha\beta}{\pi}}\ln⁡{P_{LOS}(1)}\sigma^2$ and $\eta = \alpha +2\sqrt{\frac{\alpha\beta}{\pi}}\mu+\ln⁡{P_{LOS}(1)}\times2\frac{\alpha\beta}{\pi}\sigma^2$. When $ \Gamma \geq  2
\sigma $, $\int_0^\infty \frac{1}{\sqrt{2\pi}\sigma }e^{-\frac{(l-\Gamma )^2}{2\sigma^2 }} dl >0.97$. Hence, the approximation of the expression could be
\begin{align}
  P_{LOS}  &\approx  P_0 +(1-P_0)P_{LOS}(1)^{\eta}.
\end{align}
Otherwise, when $\Gamma < 2\sigma$ , $P_{LOS}(1)< e^{-\frac{(\mu-2\sigma)\sqrt{\alpha \beta\pi}}{2\alpha\beta\sigma^2}}\ll 1$. The $P_{LOS}$ can be approximated as $P_{LOS} \approx P_0$.

Fig. \ref{fig:fig LOS} depicts the LOS probability based on our derived expression in difference urban environments, and the parameters $\alpha, \beta, \gamma$ for building deployment could be found in \cite{Holis:2008jh}. It is obvious that the LOS probability degrades when there are denser buildings in the urban scenario.

\section{Simulation and analysis}

The Wireless Insite (WI) software is used to model AA channel in a city like Ottawa \cite{Baek:2015cz}, to study the height effect on the LOS probability and the life distance of LOS/NLOS state. There are more than 100 tall buildings in an area of $600$ m $\times 1000 $ m. Most of the building heights are between $20$ m to $40$ m  and the highest one is $51$ m. Fig. 5 depicts the simulation scenario, the five red points represent the Tx location, and black points are the locations of Rx. By changing the heights of the Tx and Rx, the LOS state and NLOS state in different heights and locations are computed. The distance between two adjacent receiving points is $\Delta d = 2$ m, and the specific parameters are listed in Table \ref{tab2}.

\begin{table}
  \renewcommand\arraystretch{1.5}
    \centering
      \caption{\label{tab2} Simulation parameters}
      \begin{tabular}{ p{3cm} p{3cm}}
      \toprule
      \textbf{Parameter} & \textbf{value} \\
      \midrule
      \textbf{$\alpha$}  & 0.37 \\
      \textbf{$\beta$}  & 188   \\
      \textbf{$\lambda$} & 13.3 \\
      \textbf{TX height / m}  & 10, 15, 20, 25, 30, 50 \\
      \textbf{RX height / m} & 2, 10, 20, 30, 40, 50\\
      \textbf{number of Tx } & 5\\
      \textbf{number of Rx } & 3643\\

      \bottomrule
      \end{tabular}%
  \end{table}%

  \begin{figure}\centering
  \includegraphics[width=2in]{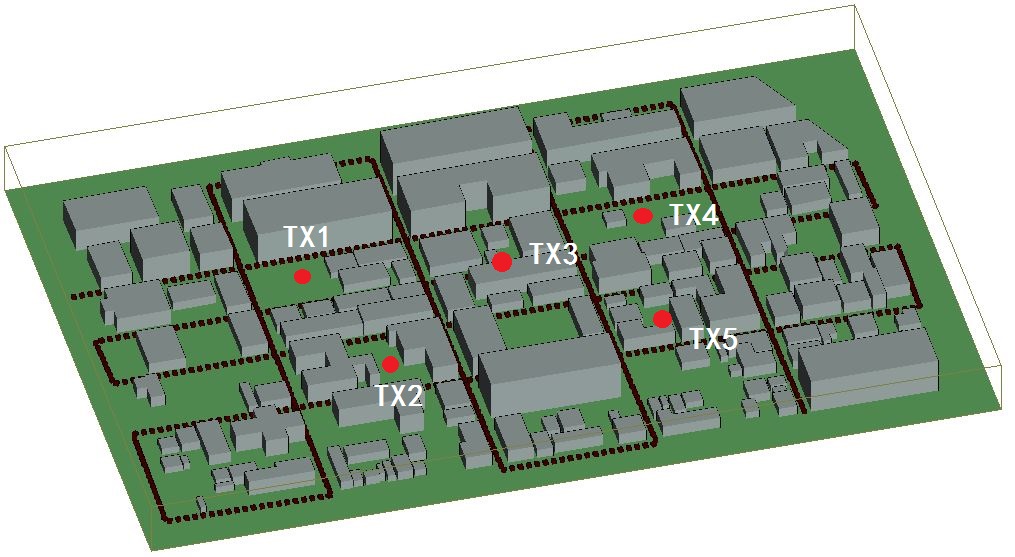}
  \vspace{-3mm}
  \caption{ Simulation scenario for AA channel in Ottawa.}
  \label{fig:fig4}
  \vspace{-3mm}
\end{figure}
\subsection{LOS probability analysis}

\begin{figure}\centering
  \includegraphics[width=2in]{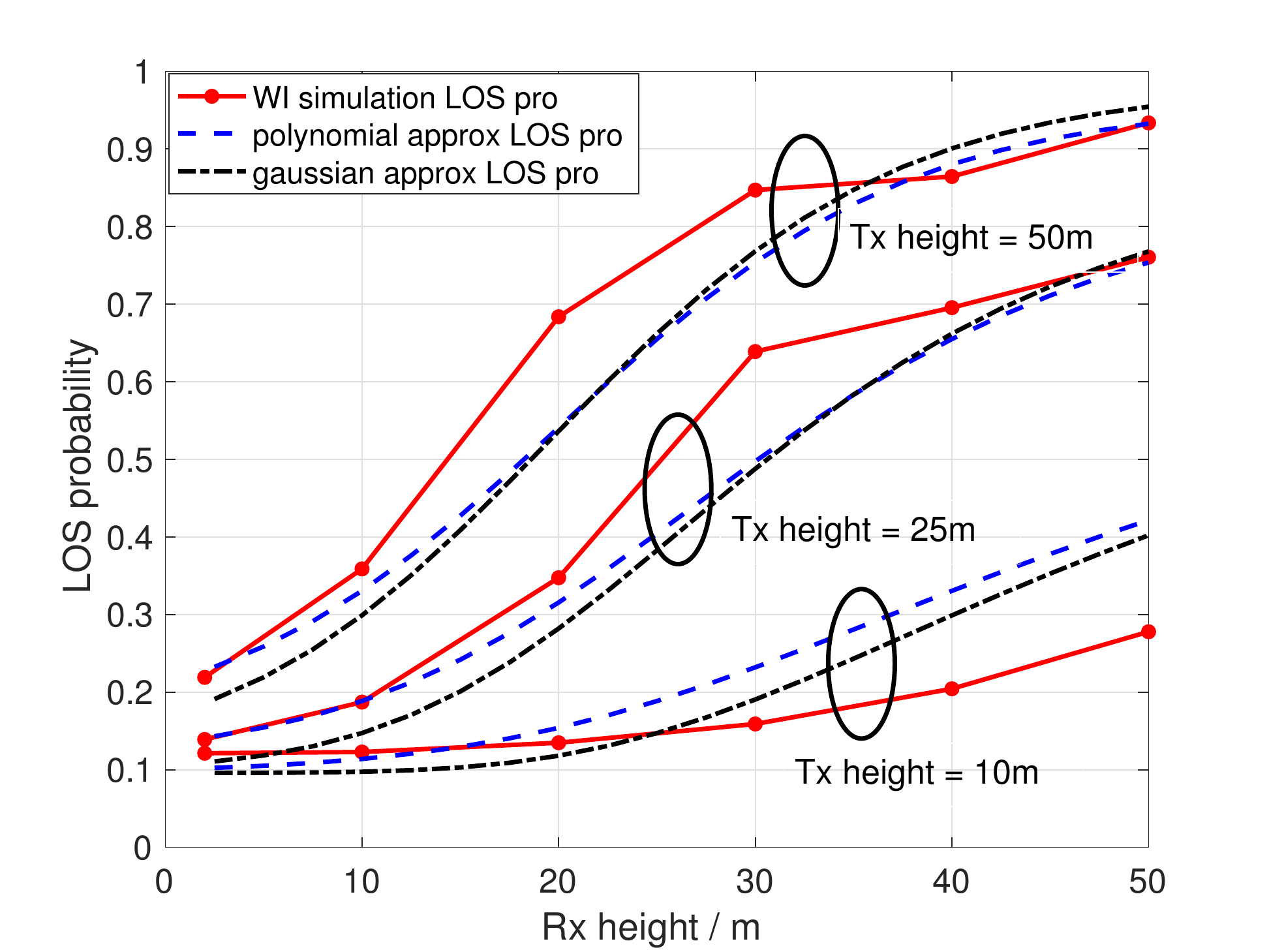}
  \vspace{-3mm}
  \caption{LOS probability at different TX and RX heights.}
  \label{fig:fig3}
  \vspace{-3mm}
\end{figure}

The polynomial approximation method (17) and Gaussian approximation method (18) are used to compute the PDF of $l$, so there are two theoretical results. Fig. \ref{fig:fig3} depicts the results of LOS probability at different TX and RX heights. In most cases, the theoretical results of LOS probability match the ray-tracing simulations well. But when the heights of Tx and Rx are $10$ m and $50$ m respectively, the estimated error could be $0.1$. The reason is that the buildings around the Tx are relatively dense in our simulation, and the LOS probability is smaller than the theoretical result when the height of Tx is low. When the heights of Tx and Rx are $25$ m and $30$ m respectively, the simulated data is larger than theoretical result, since the heights of buildings around Tx are lower than others in the simulation scenario.

\subsection{Markov model verification}

\begin{figure}\centering
    \includegraphics[width=2in]{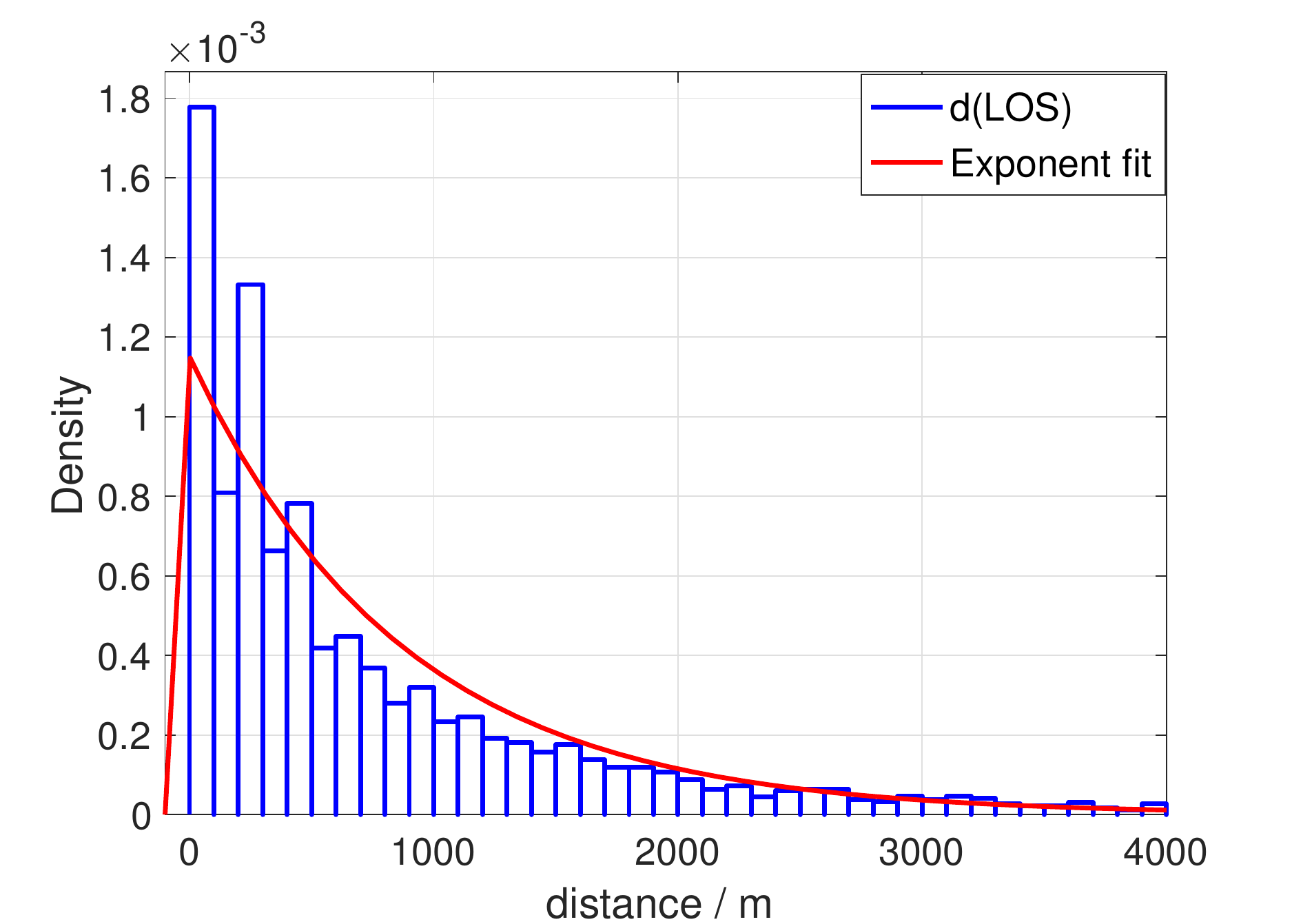}
    \vspace{-3mm}
    \caption{The distribution of life distance in LOS state, Tx height = 20 m, Rx height = 30 m.}
    \label{fig:fig7}
    \vspace{-3mm}
  \end{figure}

\begin{figure}\centering
  \includegraphics[width=2in]{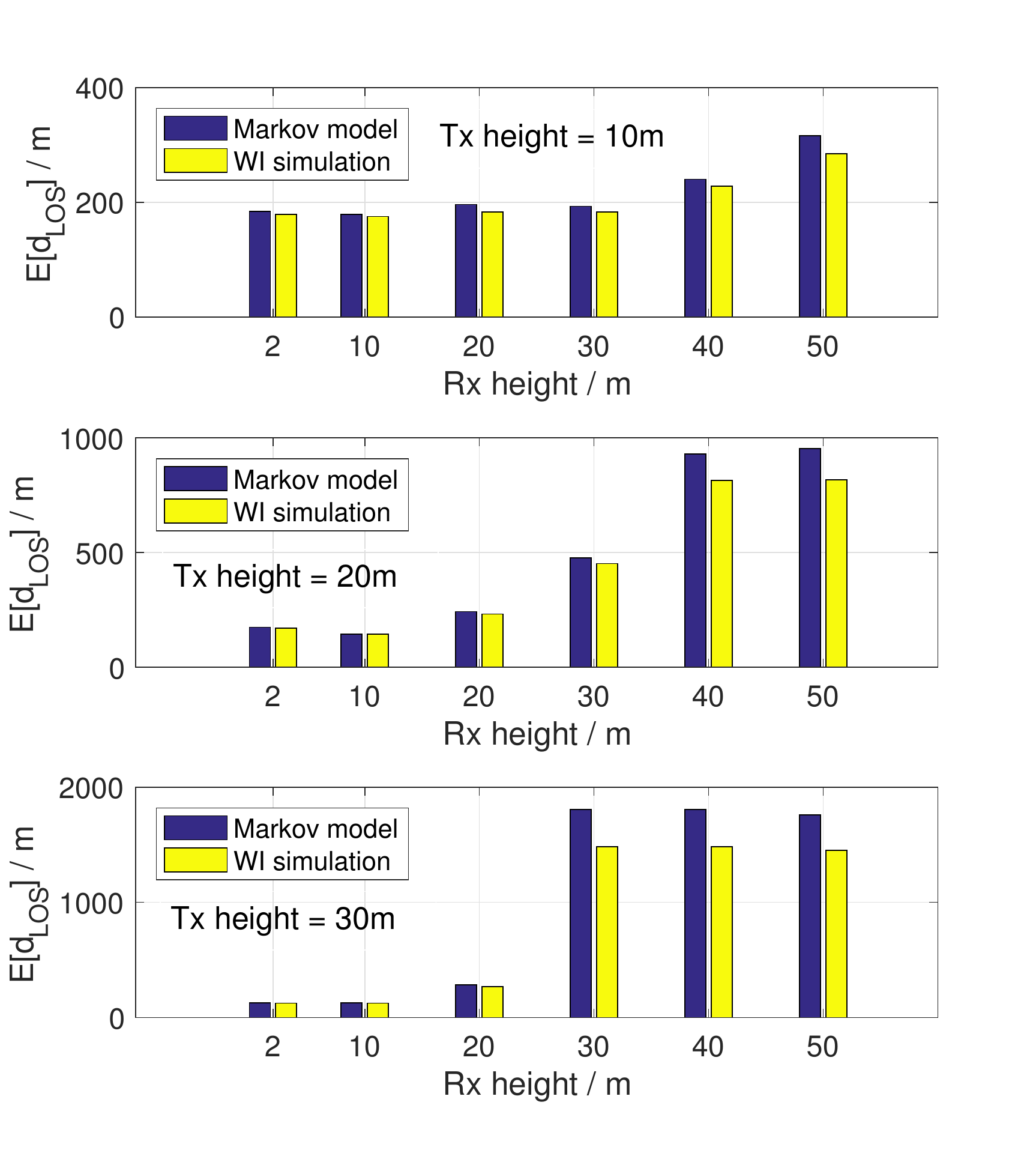}
  \vspace{-3mm}
  \caption{Life distance in LOS state at different heights.}
  \label{fig:fig8}
  \vspace{-3mm}
\end{figure}

As depicted in Fig. \ref{fig:fig7}, the PDF of $d_{LOS}$ matches the exponent distribution well, which verifies that the probability of $d_{LOS}$ can be expressed as a first-order Markov model. Fig. \ref{fig:fig8} shows the life distance in LOS state at different heights, and the theoretical results as the proposed Markov model match well with the simulation result. Moreover, the expectation of LOS distance increases with an increase in the Tx and Rx heights, because the LOS state can be maintained for a longer distance at a higher height.

\subsection{Transition rate}

\begin{figure}\centering
    \includegraphics[width=2in]{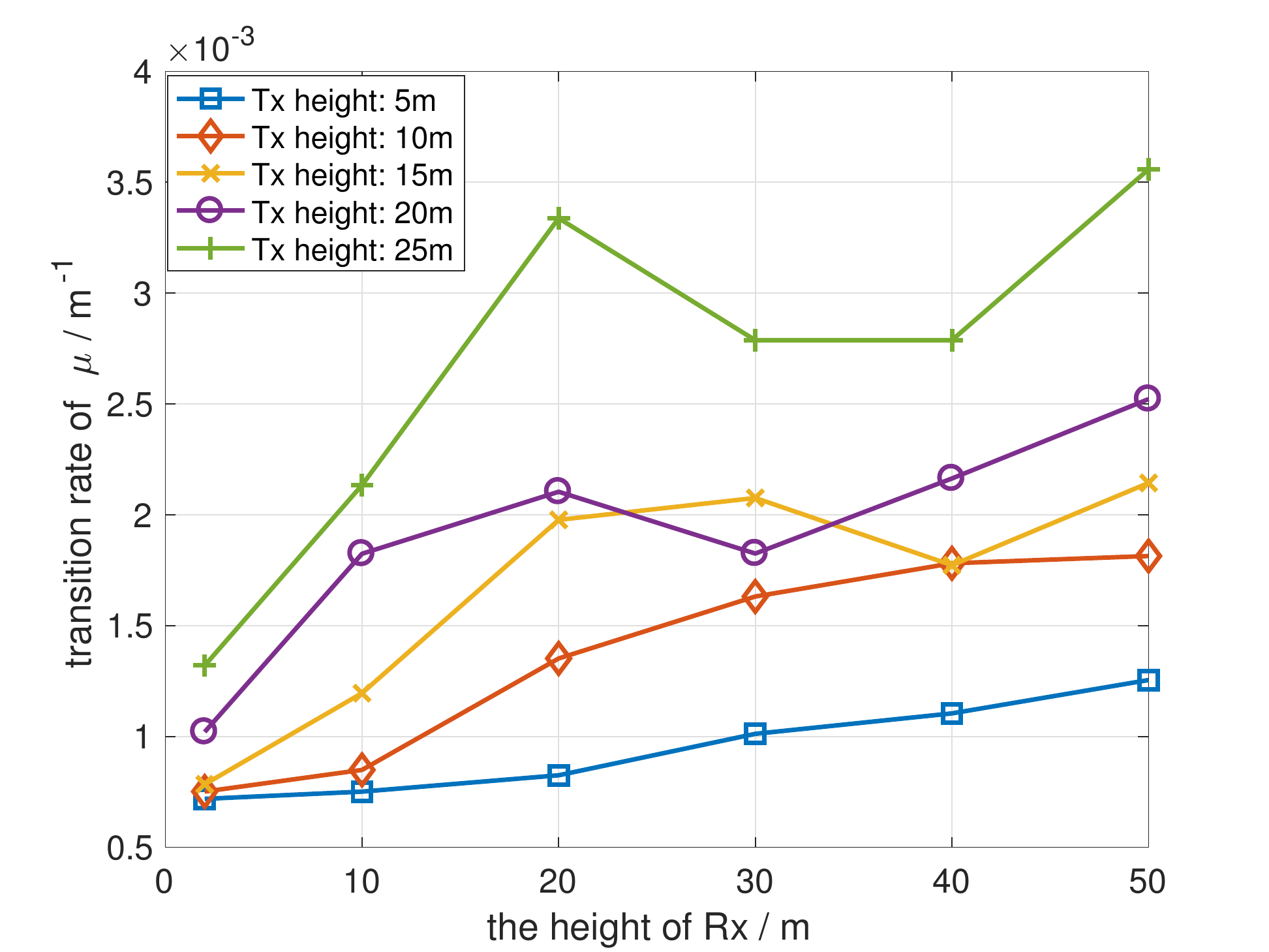}
    \vspace{-3mm}
    \caption{Transition rate $\mu$ (NLOS $\rightarrow$ LOS) at different heights.}
    \label{fig:fig5}
    \vspace{-3mm}
  \end{figure}

As depicted in Fig. \ref{fig:fig5}, when the heights of Tx and Rx are low, the transition rate $\mu$ is relatively small. As the heights of Tx and Rx increase, the transition rate $\mu$ increases gradually. The reason is that all communication links are almost in NLOS state when Tx and Rx heights are low, so that the probability from LOS state to NLOS state approaches 0. However, the LOS probability increases with an increase in TX and RX heights, accordingly the transition probability from NLOS to LOS state increases.

\begin{figure}\centering
    \includegraphics[width=2in]{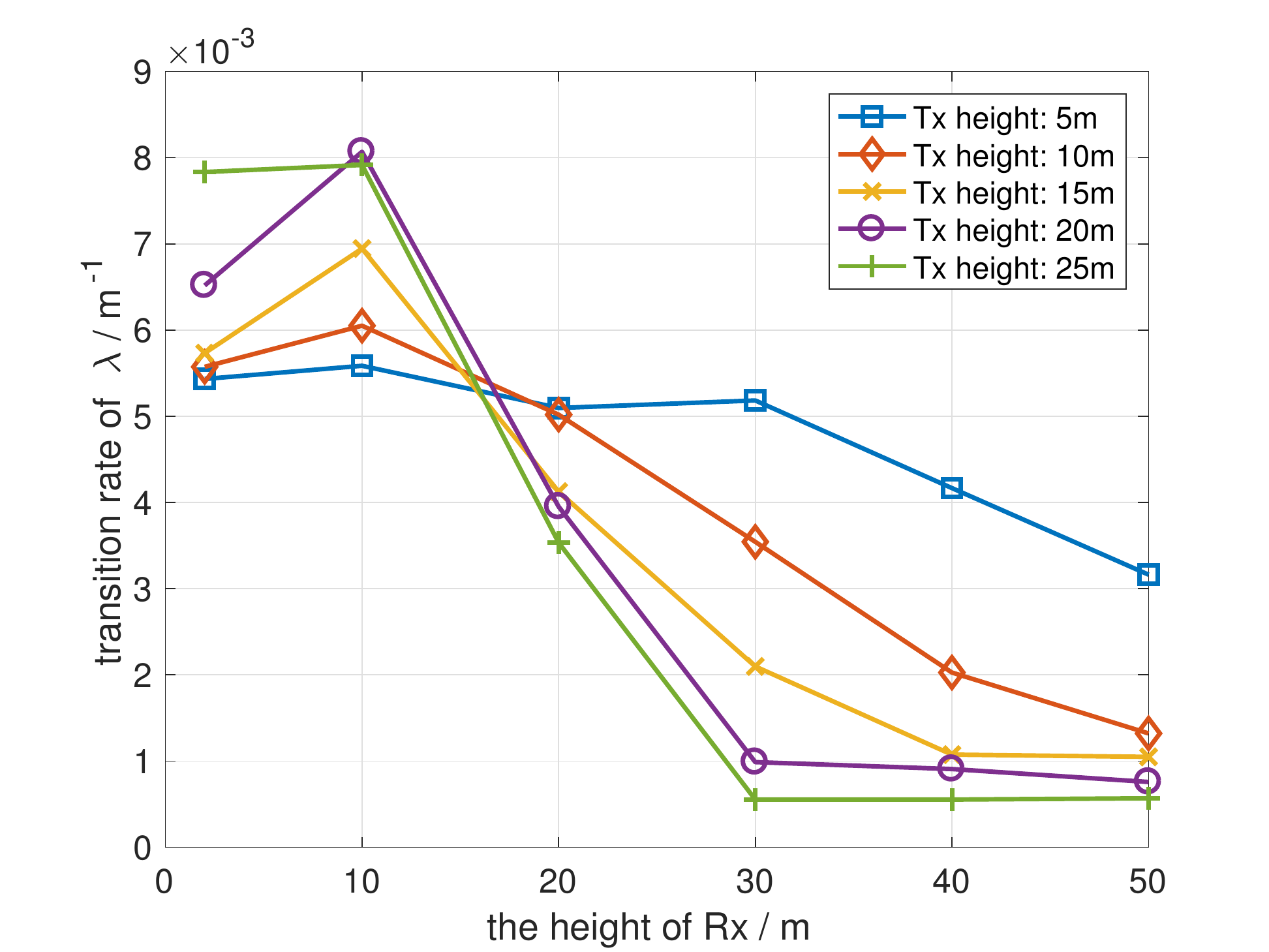}
    \vspace{-3mm}
    \caption{Transition rate $\lambda$ (LOS $\rightarrow$ NLOS) at different heights.}
    \label{fig:fig6}
    \vspace{-3mm}
  \end{figure}

As depicted in Fig. \ref{fig:fig6}, the transition rate $\lambda$ degrades with an increase in TX height when the RX is at a low altitude. But the increase of Tx height improves the $\lambda$ when the RX is at a high altitude. Because all communication links are almost in NLOS state when Tx and Rx heights are very low, and the increase in Tx height results in an increasing probability from LOS state to NLOS state. However, most of the links are in LOS state with Tx and Rx at a high altitude, and the decrease of TX heights degrades the LOS probability, hence improving the transition probability from LOS state to NLOS state.

\section{Conclusion}
In order to analyze the channel model for UAVs network communication in urban environments, we analyzed the effect of different TX and RX heights on the LOS probability. The expression of the LOS probability associated with heights of Tx and Rx is derived. And since the location of the UAV during the flight will change, a distance continuous Markov process are proposed to express the time-varying characteristics in UAV-UAV channel modeling. Based on the simulation data, the transition rate between the LOS/NLOS states are derived with their relation with the heights of UAVs.

\section{Acknowledgment}
The research presented in this paper has been kindly funded by the projects as follows, National S\&T Major Project (2017ZX03001011), National Natural Science Foundation of China (61631013), Foundation for Innovative Research Groups of the National Natural Science Foundation of China (61621091),  Tsinghua-Qualcomm Joint Project, Future Mobile Communication Network Infrastructure Virtualization and Cloud Platform (2016ZH02-3).

\end{document}